\documentstyle[12pt,aaspp4]{article}
\lefthead{Bureau \& Athanassoula}
\righthead{Bar Diagnostics in Edge-On Spiral Galaxies}
\slugcomment{Accepted for publication in The Astrophysical Journal}
\begin{document}
%
%
\newcommand{\kms}{~km~s$^{-1}$ }
\newcommand{\msun}{$M_{\mbox{\scriptsize \sun}}$ }
%
%
\title{Bar Diagnostics in Edge-On Spiral Galaxies. I. The Periodic Orbits Approach.}
\author{M.\ Bureau\altaffilmark{1}}
\affil{Mount Stromlo and Siding Spring Observatories, Institute of Advanced
Studies, The Australian National University, Private Bag, Weston Creek P.O.,
ACT~2611, Australia}
\altaffiltext{1}{Now at Sterrewacht Leiden, Postbus~9513, 2300~RA Leiden, The Netherlands}
\and
\author{E.\ Athanassoula}
\affil{Observatoire de Marseille, 2 Place Le Verrier, F-13248 Marseille Cedex
4, France}
\begin{abstract}
We develop diagnostics to detect the presence and orientation of a bar in an
edge-on disk, using its kinematical signature in the position-velocity diagram
(PVD) of a spiral galaxy observed edge-on. Using a well-studied barred spiral
galaxy mass model, we briefly review the orbital properties of two-dimensional
non-axisymmetric disks and identify the main families of periodic orbits. We
use those families as building blocks to model real galaxies and calculate the
PVDs obtained for various realistic combinations of periodic orbit families
and for a number of viewing angles with respect to the bar. We show that the
global structure of the PVD is a reliable bar diagnostic in edge-on
disks. Specifically, the presence of a gap between the signatures of the
families of periodic orbits in the PVD follows directly from the
non-homogeneous distribution of the orbits in a barred galaxy. Similarly,
material in the two so-called forbidden quadrants of the PVD results from the
elongated shape of the orbits. We show how the shape of the signatures of the
dominant $x_1$ and $x_2$ families of periodic orbits in the PVD can be used
efficiently to determine the viewing angle with respect to the bar and, to a
lesser extent, to constrain the mass distribution of an observed galaxy. We
also address the limitations of the models when interpreting observational
data.
\end{abstract}
\section{Introduction\label{sec:introduction}}
\nopagebreak
The classification of spiral galaxies along the Hubble sequence (Sandage
\markcite{s61}1961) is difficult for highly inclined systems. The tightness of
the spiral arms and the degree to which they are resolved into stars and
\ion{H}{2} regions are useless criteria when dealing with edge-on
galaxies. Only one main criterion remains: the relative importance of the
bulge with respect to the disk. The problem is more acute when it comes to
determining if a galaxy is barred, as there is no easy way to identify a bar
in an edge-on system. The presence of a plateau in the light distribution of a
galaxy (typically the light profile along the major axis) is often taken to
indicate the presence of a bar (e.g.\ de Carvalho \& da Costa
\markcite{dd87}1987; Hamabe \& Wakamatsu \markcite{hw89}1989). However, this
method has two serious shortcomings: axisymmetric features might be mistaken
for a bar (e.g.\ a lens would probably produce a very similar effect) and
end-on bars are likely to be missed (their plateaus would be both short and,
in early type barred galaxies, superposed on the steep light profile of the
bulge). The studies of the galaxy NGC~4762 by Burstein (\markcite{b79a}1979a,
\markcite{b79b}1979b), Tsikoudi \markcite{t80}(1980), Wakamatsu \& Hamabe
\markcite{wh84}(1984), and Wozniak \markcite{w94}(1994) illustrate the
uncertainties resulting from using such a method. It is clear that a
photometric or morphological identification of bars in edge-on spiral galaxies
is problematic and unsatisfactory.

Kuijken \& Merrifield \markcite{km95}(1995) (see also Merrifield
\markcite{m96}1996) were the first to demonstrate that a kinematical
identification of bars in external edge-on spiral galaxies was possible. They
calculated the projection of periodic orbits in a barred galaxy model for
various line-of-sights and showed that an edge-on barred disk produces
characteristic double-peaked line-of-sight velocity distributions which would
not occur in an axisymmetric disk. Equivalent methods have been used for many
years in Galactic studies (e.g.\ Peters \markcite{p75}1975; Mulder \& Liem
\markcite{ml86}1986; Binney et al.\ \markcite{bgsbu91}1991; Wada et al.\
\markcite{wthh94}1994; and more recently Weiner \& Sellwood
\markcite{ws95}1995; Beaulieu \markcite{b96}1996; Sevenster et al.\
\markcite{ssvf97}1997; Fux \markcite{f97a}1997a,\markcite{f97b}b), since the
PVDs of external galaxies are analogous to the longitude-velocity diagrams of
the aforementioned studies.

In this paper, we aim to develop bar diagnostics using the PVDs of edge-on
spiral galaxies in the same spirit as Kuijken \& Merrifield
\markcite{km95}(1995). We will, however, study the signature of each family of
periodic orbits separately (before joining them to obtain a global picture)
and examine how it depends on the viewing angle.  We use a well-studied mass
model, a well-defined method to populate the periodic orbits, and we explore a
large number of periodic orbit families.  Our results should be used as a {\em
guide} to interpret observations of the stellar and/or gaseous kinematics in
edge-on spiral galaxies. While the gas streamlines can be approximated by
periodic orbits, the presence of shocks will modify this behaviour
significantly. Also, the collisionless stellar component is not confined to
periodic or regular (quasi-periodic) orbits and there could be a
non-negligible fraction of stars on irregular orbits. Athanassoula \& Bureau
(\markcite{ab99a}1999a, hereafter Paper~II) and Athanassoula \& Bureau
(\markcite{ab99b}1999b, hereafter Paper~III) will provide bar diagnostics
similar to those developed here but using, respectively, hydrodynamical and
$N$-body simulations.

The identification of bars in edge-on spiral galaxies is not a goal in itself
but rather a tool allowing us to deepen our understanding of bars. The
particular line-of-sight to edge-on systems allows us to get a view of the
kinematics of the {\em entire} symmetry plane of the disk in one single
observation (assuming the disk is transparent) and provides a unique way of
studying the dynamics of the disk globally. More importantly, such a
diagnostic represents a unique opportunity to study the vertical structure of
bars, of which very little is known observationally. Three-dimensional
$N$-body simulations have shown that bars tend to buckle soon after their
formation and settle with an increased thickness and vertical velocity
dispersion, appearing boxy or peanut-shaped when viewed edge-on (e.g.\ Combes
\& Sanders \markcite{cs81}1981; Combes et al.\ \markcite{cdfp90}1990; Raha et
al.\ \markcite{rsjk91}1991). Beside the clues provided by the Galaxy (e.g.\
Blitz \& Spergel \markcite{bs91}1991; Binney et al.\ \markcite{bgsbu91}1991;
Weiland et al.\ \markcite{wetal94}1994), little observational data exist to
directly test this hypothesis. In fact, the vertical light distribution of a
bar has never been measured. Kuijken \& Merrifield \markcite{km95}(1995) and
Bureau \& Freeman \markcite{bf97}(1997) are the only ones to have actively
searched for the kinematical signature of large scale bars in
boxy/peanut-shaped bulges. Although their results seem to support the scenario
described above, only eight galaxies have been studied so far. A similar study
of a sample of over thirty galaxies, most of which have a boxy or
peanut-shaped bulge, will appear in Bureau \& Freeman
\markcite{bf99}(1999). The development of better bar diagnostics and the
search for bars in edge-on systems are the keys to a better understanding of
the vertical structure of bars. This series of papers aims to fulfill the
first of those needs; Bureau \& Freeman \markcite{bf99}(1999) will address the
latter.

In \S~\ref{sec:models}, we describe the mass model used throughout this paper
and detail the methods adopted to calculate and populate periodic orbits. 
The orbital properties of the mass model are described in \S~\ref{sec:periodfam}. 
In \S~\ref{sec:bardiag}, we describe the PVDs of edge-on spirals and develop
kinematical bar diagnostics based on the properties of prototypical barred
models with and without inner Lindblad resonances. We also generalise those
diagnostics to a large range of models. The limitations of the models for
interpreting spectroscopic observations are discussed in
\S~\ref{sec:discussion}. We summarise our results and conclude briefly in
\S~\ref{sec:conclusions}. 
\section{Models\label{sec:models}}
\nopagebreak
\subsection{Density Distribution\label{sec:density}}
\nopagebreak
The mass model used in this paper and in \markcite{ab99}Paper~II is the same
as that used by Athanassoula (\markcite{a92a}1992a,\markcite{a92b}b, hereafter
A92a, A92b); the results from all papers can therefore be directly compared
and are complementary. We briefly review the main characteristics of the mass
model here and refer the reader to \markcite{a92a}A92a for more discussion of
its properties.

The mass model has four free parameters which define the density
distribution. The bar is represented by a Ferrers spheroid (Ferrers
\markcite{f77}1877) with density
\begin{equation}
\label{eq:bar}
\rho(x,y,z)=\left\{ \begin{array}{ll}
                    \rho_0(1-g^2)^n & \mbox{for } g<1 \\
                    0 & \mbox{for } g\geq1,
                    \end{array} \right.
\end{equation}
where $g^2=x^2/a^2+(y^2+z^2)/b^2$, $a$ and $b$ are the semi-major and
semi-minor axes of the bar ($a>b$), $\rho_0$ is its central density, and
$(x,y,z)$ are the coordinates in the frame corotating with the bar.  We will
consider both homogeneous ($n=0$) and inhomogeneous models; the latter with
$n=1$. The semi-major axis is, as in A92a, fixed at 5~kpc, but, contrary to
A92a, the major axis of the bar is along the $x$-axis. We have thus so far
introduced two free parameters: the bar axial ratio $a/b$ (which fixes $b$)
and the quadrupole moment of the bar $Q_m$ (which fixes
$\rho_0$). \markcite{a92a}A92a shows how the central density, axial ratio,
quadrupole moment, and mass of the bar are related. The pattern speed of the
bar ($\Omega_p$), or equivalently the distance from the center to the
Lagrangian points $L_1$ and $L_2$ ($r_L$), constitutes a third free parameter.

The bar model described has often been used in the past and is well studied
both in the context of orbital studies (e.g.\ Athanassoula et al.\
\markcite{abmp83}1983; Papayannopoulos \& Petrou \markcite{pp83}1983; Teuben
\& Sanders \markcite{ts85}1985) and of hydrodynamical simulations (e.g.\
Sanders \& Tubbs \markcite{1980}1980; Schwarz \markcite{s85}1985). The main
deficiencies of this density distribution are that the shape and axial ratio of
the bar are independent of radius, and that the isodensities are necessarily
ellipses.

The density distribution we use has two axisymmetric components which,
when combined together, produce a rotation curve rising relatively rapidly in
the inner parts and flat in the outer parts. The first component is a
Kuzmin/Toomre disk of surface density 
\begin{equation}
\label{eq:disk}
\sigma(r)=\frac{V_0^2}{2\pi Gr_0}(1+r^2/r_0^2)^{-3/2}
\end{equation}
(Kuzmin \markcite{k56}1956; Toomre \markcite{t63}1963), where $V_0$ and $r_0$
are fixed to yield a maximum disk circular velocity of 164.2\kms at 20~kpc. 
The second axisymmetric component is a central bulge-like spherical density
distribution given by 
\begin{equation}
\label{eq:bulge}
\rho(R)=\rho_b(1+R^2/R_b^2)^{-3/2},
\end{equation}
where $\rho_b$ is the bulge central density and $R_b$ its scalelength. The
fourth free parameter of the mass model is the central concentration,
$\rho_c=\rho_0+\rho_b$ (which fixes $\rho_b$). The bulge scalelength is
determined by imposing a fixed total mass within 10~kpc. 

The models are therefore parametrised by an index $n$ ($n=0$ or 1) and by four
free parameters: the bar axial ratio $a/b$, the quadrupole moment of the bar
$Q_m$, the Lagrangian radius $r_L$, and the central concentration $\rho_c$. It
should be noted that while the quadrupole moment of the bar affects all
Fourier components of the potential equally, this is not the case for the
axial ratio. The bar pattern speed and central concentration mainly affect the
existence and position of the resonances. The models considered are those of
\markcite{a92a}A92a (see her Table~1). We will also use her units:
$10^6$~\msun for masses, kpc for lengths, and \kms for velocities.  Based on a
comparison with an observational sample (rotation curves, resonances
positions, and Fourier components), \markcite{a92a}A92a showed that these
models are a fair representation of early type barred galaxies.
\subsection{Periodic Orbits Calculations\label{sec:orbitscal}}
\nopagebreak
The periodic orbits allowed by a model are found using the shooting
method. Through this paper, we will only consider orbits in the plane of the
disk ($z=0$). For a given position along the $y$-axis ($x=0$) and an initial
velocity parallel to the $x$-axis ($\dot{y}=0$), we follow a trial orbit for
half a turn in the reference frame of the bar. Other trial orbits with the
same initial position but slightly different initial velocities allow
iterative convergence to an orbit which ``closes'' after half a
revolution. The orbits are integrated using a fourth-order Runge-Kutta method
and the Newton-Raphson method is used to converge to the right initial
velocity (Press et al.\ \markcite{pftv86}1986). By then moving the initial
position along the $y$-axis, it is possible to delineate a family of periodic
orbits. Here, we use a constant increment along the $y$-axis ($\Delta
y=0.01$~kpc for all families). All periodic orbits found in this way are
symmetric with respect to the minor axis of the bar. It should be noted that
it is possible to have more than one periodic orbit at a given position along
the $y$-axis (with different initial velocities $\dot{x}$).

In the limit of negligible pressure, gas streamlines coincide with periodic
orbits. However, contrary to periodic orbits, gas streamlines can not
intersect. Thus, because we are mainly interested in studying the gaseous
dynamics of barred spiral galaxies, we are not interested in periodic orbits
that self-intersect or possess loops. We have therefore searched and
identified only direct singly periodic non-self-intersecting orbits, which may
best represent the gas flow. This constraint limits the extent of the periodic
orbit families we have studied.

Periodic orbits can be regarded as galactic building blocks, but it is
non-trivial to determine how best to use them to represent the gas
distribution in a real galaxy. For stellar systems, Schwarzschild
\markcite{s79}(1979) proposed a method where a linear combination (with
non-negative weights) of orbits is used to reproduce the original mass
distribution (yielding a self-consistent model). Here, we simply consider all
periodic orbits from certain families to be populated equally. Whenever we
plot orbits, we plot an equal number of timesteps (an equal number of
``points'') for all orbits, independent of the period. Since we use a constant
increment along the $y$-axis between the orbits of a given family, the
resulting surface density along that axis is inversely proportional to the
distance from the center (this would be true everywhere if the orbits were
self-similar). This procedure will be used whenever we plot orbits. One shortcoming
of this method is that, although we have only selected individual periodic
orbits which do not self-intersect and do not possess loops, orbits from a
given family or from different families of periodic orbits can intersect. Such
situations could not occur in the case of gas.
\section{Periodic Orbit Families\label{sec:periodfam}}
\nopagebreak
A detailed study of the periodic orbit families located within corotation in
our models was carried out by \markcite{a92a}A92a. In this paper, we will
extend her study to the outer parts of the models (outside corotation) and
draw heavily on her conclusions to explain the behaviour observed in the inner
parts. For a more general description of the orbital structure and dynamics of
barred spiral galaxies, we refer the reader to the excellent reviews by
Contopoulos \& Grosb\o l \markcite{cg89}(1989) and Sellwood \& Wilkinson
\markcite{sw93}(1993). In this section, we will describe the main properties
of the basic families of periodic orbits present in the models. We will focus
on two inhomogeneous bar models which are prototypes of models with and
without inner Lindblad resonances (ILRs). They are, respectively: model 001
($a/b=2.5$, $Q_m=4.5\times10^4$, $r_L=6.0$, $\rho_c=2.4\times10^4$) and model
086 ($a/b=5.0$, $Q_m=4.5\times10^4$, $r_L=6.0$, $\rho_c=2.4\times10^4$). Using
the results of \markcite{a92a}A92a, it is easy to extend the conclusions drawn
from models 001 and 086 to most other models.
\placefigure{fig:chardiag}
\placefigure{fig:a/bvxvy}

Figure~\ref{fig:chardiag} shows the characteristic diagrams for models 001 and
086. For all calculated periodic orbits, they show the Jacobi integral
($E_J=E-\vec{\Omega_p}\cdot\vec{J}$) of the orbit as a function of the
position where the orbit intersects the $y$-axis. The Jacobi integral
represents the energy in the rotating frame of the bar, and is the only
combination of the energy and angular momentum which is conserved (neither
being conserved separately in a rotating non-axisymmetric potential). All the
major periodic orbit families are present. More exist, especially higher order
resonance families near corotation, but they are probably unimportant for
understanding the gas flow. Figure~3 of \markcite{a92a}A92a shows examples of
periodic orbits from the main families in model 001 (see Sellwood \& Wilkinson
\markcite{sw93}1993 for families outside corotation, although they use a
slightly different potential). The most important families inside corotation
are the $x_1$ and $x_2$. The $x_1$ orbits are elongated parallel to the bar
and are generally thought to support it (see, e.g., Contopoulos
\markcite{c80}1980). The $x_2$ (and $x_3$) orbits are elongated perpendicular
to the bar and only occur inside the ILRs. Some properties of the $x_1$ and
$x_2$ periodic orbits which will be useful in the next sections are summarised
in Figure~\ref{fig:a/bvxvy}. We do not consider the retrograde $x_4$ family
here. The inner 4:1 family (four radial oscillations during one revolution)
may be important for the structure of rectangular bars. Outside corotation,
the dominant periodic orbit families are the $x_1^{\prime}$ and outer 2:1,
corresponding to the ``$x_i$'' families inside corotation. The $x_1^{\prime}$
orbits are elongated parallel to the bar and located outside the outer
Lindblad resonance (OLR). The outer 2:1 orbits are perpendicular the bar and
located between corotation and the OLR. The short period orbits (SPO) and long
period orbits (LPO) are located around the (stable) Lagrange points $L_4$ and
$L_5$ on the minor axis of the bar.
\placefigure{fig:freq}

Figure~\ref{fig:freq} shows the main precession frequencies for models 001 and
086, obtained by azimuthally averaging the mass distribution. The major
resonances are easily identified: ILRs ($\Omega_p=\Omega-\kappa/2$), inner
ultra-harmonic resonance (IUHR; $\Omega_p=\Omega-\kappa/4$), corotation
($\Omega_p=\Omega$), and OLR ($\Omega_p=\Omega+\kappa/2$). Defined this
way, the presence of ILRs is not sufficient to guarantee the existence of the
$x_2$ family. Contopoulos \& Papayannopoulos \markcite{cp80}(1980) showed that
the $x_2$ family disappears for strong bars. For our mass model,
\markcite{a92a}A92a showed that the $x_2$ orbits are absent for small
Lagrangian radii $r_L$, low central concentrations $\rho_c$, large bar axial
ratios $a/b$, and for large quadrupole moments $Q_m$ (see Figs.~6 and 7 of
\markcite{a92a}A92a). In particular, despite the presence of ILRs in model
086, no $x_2$ orbit exists. It is thus necessary to extend the classical
definition of an ILR to the strong bar case. van Albada \& Sanders
\markcite{vs82}(1982) and \markcite{a92a}A92a propose that the existence of
ILRs can be tied with the existence of the $x_2$ periodic orbit family and the
position of the ILRs assimilated with the minimum and maximum of the $x_2$
characteristic curve in the characteristic diagram (of course, there might be
only one ILR). We will use this definition of the existence of ILRs in this
paper, which explains why model 086, despite having two ILRs in the classical
sense, is considered a ``no-ILRs'' model. Similarly, we will assimilate the
position of the IUHR with the maximum of the $x_1$ characteristic curve before
the 4:1 gap in the characteristic diagram (\markcite{a92a}A92a).
\section{Bar Diagnostics\label{sec:bardiag}}
\nopagebreak
\subsection{Detecting Edge-On Bars\label{sec:edgeon}}
\nopagebreak
In the spirit of Kuijken \& Merrifield \markcite{km95}(1995) and Merrifield
\markcite{m96}(1996), our basic tool to identify bars in edge-on disk galaxies
will be PVDs. We obtain those by calculating the projected density of material
in our edge-on barred disk models as a function of line-of-sight velocity and
projected position along the major axis (for various line-of-sights). These
can then be directly compared with long-slit spectroscopy observations of
edge-on spiral galaxies (with the slit positioned along the major axis) or
with other equivalent data sets. The goal is to identify features in the PVDs
which can be unmistakably associated with the presence of a bar. We discuss
such features in the next sections.
\subsection{Model 001 (ILRs)\label{sec:001}}
\nopagebreak
\placefigure{fig:x1}
\placefigure{fig:x2}
\placefigure{fig:out2_1d}

Figures~\ref{fig:x1}, \ref{fig:x2}, and \ref{fig:out2_1d} show PVDs for,
respectively, the $x_1$, $x_2$, and outer 2:1 periodic orbit families of
model 001, which has ILRs (or, equivalently, has an $x_2$ family of periodic
orbits). Each figure presents the face-on appearance of the entire family of
orbits (with orbits equally spaced along the $y$-axis and the extent of the
family limited by gaps in the characteristic curve or the appearance of
loops in the orbits) and PVDs obtained using an edge-on projection and various
viewing angles with respect to the bar. The viewing angle $\psi$ is defined to
be 0\arcdeg\ when the bar is seen end-on and 90\arcdeg\ when the bar is seen
side-on. 

The upper left panel of Figure~\ref{fig:x1} shows that, because of the high
curvature of the $x_1$ orbits on the major axis of the bar
(\markcite{a92a}A92a) and the crowding of orbits at its ends, overdensities of
material are created which are analogous to those caused by shocks in
hydrodynamical simulations (see Sanders, Teuben, \& van Albada
\markcite{stv83}1983; Sanders \& Tubbs \markcite{st80}1980;
\markcite{a92b}A92b). As expected, very high radial velocities are present in
the PVDs when the bar is seen end-on due to streaming up and down the
bar. Conversely, the velocities are low when the bar is seen side-on because
the movement is mostly perpendicular to the line-of-sight. In the next few
paragraphs, we will analyse this effect in more detail, in order to understand
the variation of the shape of the signature of the orbits in the PVDs as a
function of the viewing angle.

In general, the trace in a PVD of a two-dimensional elongated orbit seen
edge-on can be thought of as a parallelogram. For the sake of simplicity, we
will consider here an orbit which is symmetric about two perpendicular axes
and which is centered at their origin, like the $x_1$ and $x_2$ orbits. If the
orbit is seen exactly along one of its symmetry axes, then its trace in a PVD
will be a line, both near and far sides of the orbit yielding the same radial
velocity at a given projected distance from the center. In addition, the
observed radial velocity will switch from positive to negative values at the
center (the radial velocity is null at that point). However, for all other
viewing angles, the trace of the orbit in a PVD will be strongly
parallelogram-shaped, populating the ``forbidden'' quadrants (top-right and
bottom-left quadrants of the PVDs considered here). This shape is due to the
fact that, when the line-of-sight is not parallel to an axis of symmetry of
the orbit, the near and far sides of the orbit yield different radial
velocities for a given projected distance from the center, and the position at
which the observed radial velocity switches from positive to negative values
is not the center, but rather is displaced slightly away from it. At that
position, by definition, the tangent to the orbit is perpendicular to the
line-of-sight. One only needs to look at the radial component of the velocity
along an elongated orbit to see these effects. Generally, the highest
tangential velocity occurs on the minor axis of the orbit and is parallel to
its major axis. The opposite is also generally true (but not always) for the
lowest velocity (see, e.g., Fig.~\ref{fig:a/bvxvy}). Therefore, the
parallelogram-shaped trace of an elongated orbit in a PVD is narrow but
reaches high radial velocities (with respect to the local circular velocity)
for viewing angles close (parallel) to its major axis, while it is rather
extended and reaches only relatively low radial velocities for viewing angles
close to its minor axis. The exact shape of the parallelogram in a PVD depends
primarily on the axial ratio of the orbit. For a given azimuthally averaged
radius, as the eccentricity of the orbit is increased, the velocity contrast
of the orbit (the difference between the highest and lowest tangential
velocities) also increases. The viewing angle dependence of the trace of the
orbit in a PVD is thus accentuated. At the other end of the eccentricity
range, the trace of a circular orbit in a PVD is an inclined straight line
passing through the origin and identical for all viewing angles.

The parallelogram-shaped signature of the $x_1$ orbits observed in the PVDs of
Figure~\ref{fig:x1} can be understood based on the above principles. The axial
ratio of the $x_1$ orbits generally increases with decreasing radius (except
in the very center, see Fig.~\ref{fig:a/bvxvy}a). The inner orbits will thus
reach very high radial velocities (compared to the circular velocity) at small
projected distances for small viewing angles, while they will reach only low
radial velocities at large viewing angles. On the other hand, the projected
velocities of the outer orbits will vary little with the viewing angle because
they are rounder. They will thus reach radial velocities close to the circular
velocity at large projected distances for all viewing angles. Orbits of
intermediate radii have intermediate axial ratios and thus intermediate
behaviours in the PVDs. As one moves inward in radius and in projected
distance, the locus of the maxima of the traces of successive orbits in the
PVDs will therefore increase rapidly for small viewing angles (see
Fig.~\ref{fig:a/bvxvy}b), while it will decrease for large viewing angles
(see Fig.~\ref{fig:a/bvxvy}c). This is exactly the behaviour observed in the
PVDs for the upper part of the envelope of the signature of the $x_1$ orbits
(see Fig.~\ref{fig:x1}). For orbits of very small radii, the axial ratio
actually decreases rapidly with decreasing radius (the axial ratio reaches a
maximum for orbits of minor axis length about 0.2~kpc;
Fig.~\ref{fig:a/bvxvy}a). This explains why the envelope of the signature of
the $x_1$ orbits does not increase right to the center, but drops abruptly
just before that. The ellipsoidal ``holes'' in the PVDs at intermediate
viewing angles are due to the fact that we stopped the $x_1$ periodic orbits
at the IUHR, not populating the small segments of the $x_1$ characteristic
curve existing past the 4:1 gap in the characteristic diagram (see Fig.~2 in
\markcite{a92a}A92a). The holes disappear if we include orbits at larger
Jacobi constant $E_J$ (which are also rounder).

In Figure~\ref{fig:x2}, the behaviour of the $x_2$ orbits can be contrasted
with that of the $x_1$. As expected, because the $x_2$ orbits are elongated
perpendicular to the bar (see Fig.~3 of \markcite{a92a}A92a), the highest
radial velocities are now reached when the bar is seen side-on, and the lowest
when the bar is seen end-on. The general parallelogram shape is still present,
but its nature is quite different than that of the signature of the $x_1$
orbits shown in Figure~\ref{fig:x1}. Contrary to the $x_1$ orbits, the axial
ratio of the $x_2$ orbits generally decreases with decreasing radius (up to
about 0.4~kpc, see Fig.~\ref{fig:a/bvxvy}d). The inner orbits have only a
short extent and, because they are almost circular, they do not reach high
radial velocities. Their projected velocity is close to the circular velocity
for all viewing angles. The outer orbits, on the other hand, are highly
elongated. They will thus reach only relatively low radial velocities at
``large'' projected distances for small viewing angles, and high velocities at
``large'' distances for large viewing angles (they are elongated perpendicular
to the bar). The locus of the maxima of the traces of successive orbits of
decreasing radius in the PVDs will therefore increase rapidly for small
viewing angles (see Fig.~\ref{fig:a/bvxvy}e) and decrease for large viewing
angles (see Fig.~\ref{fig:a/bvxvy}f). Indeed, this behaviour is observed in
the PVDs of Figure~\ref{fig:x2}, at least for ``large'' projected
distances. The behaviour at very small radii is dominated by the shape of the
circular velocity curve, which rises rapidly with radius.

The observed behaviours of the $x_1$ and $x_2$ orbits are qualitatively rather
similar. This might be surprising on first thought, as the $x_2$ orbits behave
very differently than the $x_1$, but one could say that the properties of the
$x_2$ orbits are ``doubly-inverted'' with respect to those of the $x_1$. The
variations of the axial ratio of the $x_1$ and $x_2$ orbits with radius are
opposite (Fig.~\ref{fig:a/bvxvy}), so the dependence of their signatures on
the viewing angle with respect to their major axes will be
opposite. Furthermore, the major axes of the $x_1$ and $x_2$ orbits are also
perpendicular to each other. This ``double inversion'' leads to the similarity
of the signatures in the PVDs. While this is true in a relative manner, it is
not true in an absolute way. The envelope of the signature of the $x_1$ orbits
reaches higher radial velocities than that of the $x_2$ orbits at small
viewing angles, and the opposite is observed at large viewing angles. The
explanation is simple: for small viewing angles, the radial velocities reached
by the $x_1$ orbits in the {\em inner parts are increased} with respect to the
circular velocity (the outer parts are unchanged), while for the $x_2$ orbits,
the radial velocities in the {\em outer parts are decreased} with respect to
the circular velocity (the inner parts are unchanged). The opposite is true at
large viewing angles.  A further difference is that in the case of the $x_1$
orbits, the center of the parallelogram-shaped signature is relatively faint
compared to its edges, while for the $x_2$ orbits, it is the center of the
parallelogram which is bright, forming a strong inverted S-shaped feature, and
the outer parts are relatively faint. The S-shape feature is not due to a
single orbit leaving such a trace in the PVDs, but rather to the crowding of
the traces of many successive orbits, which explains why it is so bright (an
effect comparable to the spiral arms created by rotating slightly similar
ellipses of increasing radii; Kalnajs \markcite{k73}1973). Furthermore,
because the axial ratio of the $x_2$ orbits increases with radius (outside
0.4~kpc), the trace of the largest orbit in the PVDs is not only the most
extended but is also the one with the widest parallelogram shape. It therefore
encompasses the traces of all the other orbits and defines largely by itself
the envelope of the signature of the $x_2$ orbits, which is then very
faint. The small ``holes'' present in the center of the PVDs at intermediate
viewing angles are due to the fact that, although the elongation of the $x_2$
orbits generally decreases inward, the $x_2$ family does not extend up to the
center (see Fig.~\ref{fig:chardiag}).

Figure~\ref{fig:out2_1d} illustrates the signature of the outer 2:1 orbits in
the PVDs. Because the orbits are almost circular, the upper part of the
envelope reaches radial velocities close to the circular velocity at large
projected distances, independent of the viewing angle. The features seen in
the signature of the outer 2:1 orbits are largely due to the ``dimples'' in
the orbits on the major axis of the bar (see Fig.~11 in Sellwood \&
Wilkinson \markcite{sw93}1993). As should be expected, the PVDs for
the $x_1^{\prime}$ orbits (not shown) are similar to that of the outer 2:1
orbits when the viewing angles are reversed (e.g.\ $67.5\arcdeg \rightarrow
22.5\arcdeg$), the major axes of the orbits being at right angles. Both
families yield a slowly-rising almost solid-body signature in the PVDs for all
viewing angles.
\placefigure{fig:4:1}

Bars in early-type spirals and in $N$-body simulations tend to be more
rectangular shape rather than ellipsoidal shape (see, e.g., Sparke \& Sellwood
\markcite{ss87}1987; Athanassoula et al.\ \markcite{amwpplb90}1990).
Interestingly, the maximum boxiness generally occurs just before the end of
the bar (Athanassoula et al.\ \markcite{amwpplb90}1990), where the $x_1$
orbits are slightly boxy and where the rectangular-shaped inner 4:1 orbits are
found (see Fig.~3 of \markcite{a92a}A92a). It is thus tempting to associate the
branch of the inner 4:1 family of periodic orbits lying outside the
characteristic curve of the $x_1$ orbits in the characteristic diagram
(Fig.~\ref{fig:chardiag}) with the rectangular shape of bars. The 4:1 gap in
model 001 is of type 2 (see Contopoulos \markcite{c88}1988;
\markcite{a92a}A92a), so the lower branch of the 4:1 characteristic is stable
and the proposed association makes sense, but this is not necessarily the case
in real galaxies. In fact, early-type galaxies seem to have 4:1 gaps of type 1
(Athanassoula \markcite{a96}1996). Figure~\ref{fig:4:1} shows the surface
density and PVDs for both the $x_1$ family of periodic orbits in model 001 and
the lower branch of the inner 4:1 family. It shows that indeed the inner 4:1
orbits can create a very rectangular surface density distribution when
combined with the $x_1$ orbits. Although the signature of the inner 4:1 orbits
in the PVDs is very peculiar and easily identifiable when taken alone, it is
superposed on the signature of the $x_1$ orbits for most viewing angles and it
is hard to disentangle the two families. However, when the bar and the inner
4:1 orbits are seen either end-on or side-on, the inner 4:1 orbits leave a
signature in the PVDs distinct from that of the $x_1$ orbits. The lower limit
of the combined envelope of the signatures of the $x_1$ and inner 4:1 orbits
is straight and only slightly inclined until it does a sharp bend at
approximately the position of the IUHR (at a slightly smaller radius when
$\psi=0\arcdeg$ and slightly larger radius when $\psi=90\arcdeg$, following
the definition of the IUHR adopted in \S~\ref{sec:periodfam}); it then rises
vertically until it joins with the upper limit of the envelope. This is easily
understandable considering the morphology of the inner 4:1 orbits. The
projected edges of the density distribution are sharpest at those viewing
angles and the line-of-sights are parallel to the approximately straight
segments of the orbits (see Fig.~\ref{fig:4:1}).

The advantage of the periodic orbits approach is that various orbital
components of a galaxy can be combined together in a multitude of ways. A
superposition of the $x_1$, $x_2$, and outer 2:1 families of periodic orbits
should give a reasonable representation of a prototypical barred galaxy with
ILRs. Indeed, in the inner parts, only three direct families exist: the $x_1$,
$x_2$, and $x_3$ (see Fig.~\ref{fig:chardiag}). The $x_1$ orbits, parallel
to the bar, are certainly present and, because the $x_3$ orbits are unstable
(e.g. Sellwood \& Wilkinson \markcite{sw93}1993), the $x_2$ orbits will
dominate over the $x_3$, if orbits perpendicular to the bar are present. In
the outer parts, we find the $x_1^{\prime}$ and outer 2:1 families. The shapes
of these orbits are almost identical to that of the two subclasses of outer
rings observed in barred spiral galaxies (Buta \& Crocker
\markcite{bc91}1991): R$_1^{\prime}$ outer rings for the outer 2:1 orbits and
R$_2^{\prime}$ for the $x_1^{\prime}$ orbits. The R$_1^{\prime}$ class is
dominant (Buta \markcite{b86}1986), which is why we have chosen the outer 2:1
family of periodic orbits.  However, the signature of the $x_1^{\prime}$
orbits in the PVDs is very similar to that of the outer 2:1 orbits (almost
identical if the viewing angles are reversed) and using one or the other does
not affect our conclusions or the nature of the bar diagnostics in the
PVDs. Both families act as a slowly rising almost solid-body component.
\placefigure{fig:all}

Figure~\ref{fig:all} shows the surface density and PVDs obtained by
superposing the $x_1$, $x_2$, LPO, and outer 2:1 families of periodic orbits
for model 001. The surface density of the $x_1$, $x_2$, and outer 2:1 families
is qualitatively similar to what is observed for barred galaxies with an outer
ring. More interesting is the amount of structure present in the PVDs,
especially in the inner parts of the model where the effects of the bar are
strongest. We find back the signatures of the $x_1$ and $x_2$ orbits already
discussed above, as well as the features which allow us to determine the
viewing angle with respect to the major axis of the bar. The large gap between
the signatures of the $x_1$ and outer 2:1 families of periodic orbits is due
to a corresponding gap in our reconstitution of the density distribution of a
prototypical barred galaxy. The only way to make this gap disappear is to
populate periodic orbit families close to corotation, but this is not obvious
using our periodic orbits approach. Beyond the 4:1 gap in the characteristic
diagram, there are higher order n:1 families, and, between consecutive 2n:1
gaps, short segments of what can still be called $x_1$ orbits (see
Fig.~\ref{fig:chardiag} and Fig.~2 in \markcite{a92a}A92a). However, as
can be seen from the figures of \markcite{a92b}A92b, the gas streamlines
continue to be ellipsoidal and elongated along the bar past the IUHR, the
extent of this region being very model dependent. Loosely speaking, one could
say that, although the gas does not follow precisely the higher order
resonance families, it follows their general form. Yet further out the gas
circulates around each of the two stable Lagrangian points $L_4$ and $L_5$,
the streamlines now being associated with the LPO periodic orbits (see
Fig.~11 in Sellwood \& Wilkinson \markcite{sw93}1993). The latter are easy
to add in our description and have also been included in
Figure~\ref{fig:all}. Their signature in the PVDs is very similar to that of
circular orbits. As expected, because of their location, the signature of the
LPO orbits falls right in the gap between the signature of the $x_1$ orbits
and that of the outer orbits. This gap is significantly reduced, but many
smaller gaps are still present because of the non-homogeneous distribution of
the orbits of the various families. Such gaps could not occur in an
axisymmetric spiral galaxy.
\subsection{Model 086 (no-ILRs)\label{sec:086}}
\nopagebreak
\placefigure{fig:086}

Despite the presence of ILRs in the classical sense (see
Fig.~\ref{fig:freq}), we consider model 086 a ``no-ILR'' model because it
does not have an $x_2$ family of periodic orbits. Its characteristic diagram
is very similar to that of model 001 (Fig.~\ref{fig:chardiag}), differing
only in the inner parts, where the $x_2$ and $x_3$ families are absent and the
$x_1$ characteristic curve displays an elbow due to the high axial ratio of
the bar (see also Fig.~4 of \markcite{a92a}A92a; Pfenniger
\markcite{p84}1984). In addition, the $x_1$ orbits possess loops for a certain
range of radii. Here, we exceptionally include those orbits to prevent the
appearance of an empty region in the $x_1$ orbits surface density
distribution.

Figure~\ref{fig:086} shows the surface density and PVDs obtained by
superposing the $x_1$ and outer 2:1 families of periodic orbits for model
086. The surface density distribution is again similar to that observed for
the gaseous component in barred spiral galaxies. In the PVDs, as expected, the
signature of the outer parts of the model has not changed, the outer 2:1
family of periodic orbits behaving again like a slowly-rising solid-body
component. In the inner parts, the obvious difference with the PVDs of model
001 (Fig.~\ref{fig:all}) is the absence of the signature of any $x_2$ orbit
(the LPO orbits have been omitted in Fig.~\ref{fig:086} for clarity). The
signature of the $x_1$ orbits has changed only slightly on a qualitative
level, the envelope still being generally parallelogram-shaped.  The main
difference with the signature of the $x_1$ orbits in model 001 is that the
envelope of the signature has more curved edges, due to the presence of orbits
with loops, and reaches more extreme radial velocities (compared to the
circular velocity), due to the higher axial ratio of the bar yielding more
eccentric orbits (see Fig.~10 in \markcite{a92a}A92a). The gap between the
signatures of the $x_1$ and outer 2:1 orbits is again due to the absence of
populated orbits near corotation in our model.
\subsection{Other Models\label{sec:others}}
\nopagebreak
Now that we understand the general structure of the PVDs produced by models
with and without ILRs, we can extend our study to investigate how the bar
diagnostics might change when the free parameters of the mass model are
varied. To do this, we borrow heavily on the results of \markcite{a92a}A92a,
who studied how the orbital structure of the mass model varies within most of
the volume of parameter space likely to be occupied by real galaxies. 
We do not expect the outer parts of the models to vary significantly since the
influence of the bar falls off rapidly with radius. The outer families of
periodic orbits will always produce slowly-rising almost solid-body signatures
in the PVDs. We will thus concentrate on understanding the behaviour of the
periodic orbits in the inner parts of the models.

The parallelogram-shaped signatures of the $x_1$ and $x_2$ periodic orbits in
the PVDs will be mainly affected by their eccentricity and extent. The results
of \markcite{a92a}A92a concerning the eccentricity of the $x_1$ orbits can be
summarised as follows (see Fig.~10 in \markcite{a92a}A92a): as the axial
ratio $a/b$ of the bar is increased, the Lagrangian radius $r_L$ increased,
the central concentration $\rho_c$ increased, and/or the quadrupole moment
$Q_m$ of the bar decreased, the eccentricity of the $x_1$ orbits is
increased. The counterintuitive behaviour of the eccentricity of the $x_1$
orbits with $Q_m$ stems from the fact that for $Q_m$ to be increased, the
bulge mass and therefore the central density of the model has to be decreased
(the total mass within a given radius being fixed), leading to a decrease in
the eccentricity of the orbits. The eccentricity of the $x_2$ orbits behaves
in the same way as that of the $x_1$ orbits except with respect to the axial
ratio of the bar (again, see Fig.~10 in \markcite{a92a}A92a). In that case,
the $x_2$ orbits become less eccentric as the bar axial ratio is
increased. Because higher eccentricity means more extreme radial velocities
compared to the circular velocity in the PVDs (very high when the orbit is
seen end-on and very low when the orbit is seen side-on), the envelopes of the
signatures of the $x_1$ and $x_2$ orbits in the PVDs should be most extreme
(in the above sense) for high bar axial ratios (except for the $x_2$ orbits),
high Lagrangian radii, high central densities, and/or low bar quadrupole
moments. This was certainly the case for model 086, which has a higher bar
axial ratio than model 001.

\markcite{a92a}A92a showed that the radial extent of the $x_1$ family is
mainly affected by the pattern speed of the bar and changes very little as the
other parameters of the mass model are varied (see Fig.~6 and 7 of
\markcite{a92a}A92a). For the $x_2$ orbits, the major factor affecting their
signature in the PVDs will be their existence or non-existence, depending on the
model considered. \markcite{a92a}A92a showed that the radial range of the
$x_2$ orbits is reduced when the bar axial ratio $a/b$ and/or quadrupole moment
$Q_m$ are increased, and when the central density $\rho_c$ and/or Lagrangian
radius $r_L$ are decreased (see Fig.~6 and 7 of \markcite{a92a}A92a). 
Furthermore, as exemplified by model 086, the $x_2$ orbits can be completely
absent for high bar axial ratios and/or quadrupole moments, and for low central
densities and/or Lagrangian radii. The presence and extent of the inverted
S-shape signature of the $x_2$ orbits in the PVDs depends therefore strongly
on the parameters of the model. 

Full sequences of PVDs as each parameter of the mass model is varied will be
provided in \markcite{ab92}Paper~II using hydrodynamical simulations. We do
not present them here using the periodic orbits approach to avoid unnecessary
repetition.
\section{Discussion\label{sec:discussion}}
\nopagebreak
The bar diagnostics we have developed in the previous sections are all based
on the use of families of periodic orbits in the equatorial plane of a barred
spiral galaxy mass model. Periodic orbits, however, are only an {\em
approximation} to the dynamical structure of either the gas or the stars in
galaxies and the PVDs presented in the previous sections should only be used
as a {\em guide} when interpreting kinematical data. The ability of the gas to
dissipate energy changes the behaviour of the gaseous component from that
predicted by the periodic orbits, particularly near shocks, occuring at the
transition regions between different orbit families and near periodic orbits
with loops. The kinematics of stars on regular orbits are relatively well
approximated by that of the periodic orbits, since the former are trapped
around the latter. On the other hand, stars on chaotic orbits give a totally
different signature, and the percentage of stars on such orbits may well be
non-negligible, particularly in strongly barred galaxies. In addition, we have
not attempted to make the models self-consistent when populating the orbit
families. We have only calculated the shape of the signature in the PVDs of
each family of periodic orbits of the models, but not the relative weights of
the families or of the orbits within them. Nevertheless, in order to assess
how much our results depend on the method adoped to populate the orbits, we
have also produced PVDs for the $x_1$ and $x_2$ periodic orbits of model 001
using equal increments of the Jacobi constant between orbits (rather than
equal $\Delta y$). As expected, the envelope of the signature of the $x_1$
orbits in the PVDs does not change but, because of the form of the $x_1$
characteristic curve (Fig.~\ref{fig:chardiag}), the central parts are much
stronger. Similarly, the signature of the $x_2$ periodic orbits changes very
little. Independent of those issues, the relative amplitude of each component
of the PVDs will also vary depending on the emission line used to measure the
kinematics, this simply because each component arises from a different part of
the galaxy where the line might be produced in a different way. For example,
the presence of shocks and/or increased star formation in the components will
lead to different emission line strengths in each of the component, and the
ratios will vary depending on the lines used.

When interpreting data based on the PVDs produced here, one has therefore to
take into account the following effects: 1) the kinematical signature observed
might be somewhat different from that calculated here because periodic orbits
are only an approximation to the gaseous or stellar dynamics in a galaxy, 2)
the relative amplitude of each component will be different from that
calculated because the building blocks approach used may not represent the
relative weights correctly, and 3) the observed relative amplitude of each
component will be different from that calculated because the intensity of a
line depends not only on the density of the emitting material but also on the
production mechanism of the line, which is not considered here. The
hydrodynamical simulations reported in \markcite{ab99}Paper~II and the
$N$-body simulations reported in \markcite{ab99b}Paper~III cover the first and
second problems. However, to remedy the third problem raised above, one would
need to consider both stellar evolution and the detailed physical conditions
in the gas.

The presence of dust can also hinder our ability to detect bars in edge-on
spiral galaxies. Because the dust in disks is mostly confined to a thin layer,
it can make a disk optically thick at optical wavelengths if the galaxy is
seen edge-on. If this is the case, there are two ways around the opacity
problem. First, it is possible to select objects which are not perfectly
edge-on. The line-of-sight then reaches the central parts of the galaxy where
the bar resides while still going through a substantial fraction of the
disk. However, if the inclination is too large, the bar diagnostics developed
here will not work, as they depend on the line-of-sight going through most of
the disk. Secondly, it is possible to use observations in a part of the
spectrum where even a dusty disk is likely to be optically thin. Long-slit
spectroscopy in the near-infrared (e.g.\ using the Br$\gamma$ line at
$K$-band) is attractive but most lines are weak in non-active galaxies and
near-infrared spectrographs with sufficient resolution for kinematical work
are still uncommon. A better option is to use line-imaging in the 21~cm
\ion{H}{1} line with a radio synthesis telescope. Even very dusty edge-on
spiral galaxies are probably optically thin at 21~cm. In addition, it is
possible to use a higher spectral resolution than available with most optical
long-slit spectrographs. However, radio synthesis observations are useful only
for large \ion{H}{1}-rich galaxies because of limited sensitivity and spatial
resolution. Using a large sample of galaxies, Bureau \& Freeman
\markcite{bf99}(1999) will address in more detail the question of dust
extinction when identifying bars in edge-on spiral galaxies.
\section{Summary and Conclusions\label{sec:conclusions}}
\nopagebreak
Our main goal in this paper was to develop kinematical bar diagnostics for
edge-on spiral galaxies. Considering a well-studied family of mass models
including a Ferrers bar, we identified the major periodic orbit families and
briefly reviewed the orbital structure in the equatorial plane of the mass
model. We considered only orbits which are direct, singly periodic, and
non-self-intersecting. Using a simple method to populate these orbits, we then
used the families of periodic orbits as building blocks to model the structure
of real galaxies.

We constructed position-velocity diagrams (PVDs) of the models using an
edge-on projection and various viewing angles with respect to the bar. We
considered mainly two models which are prototypes of models with and without
inner Lindblad resonances. The PVDs obtained show a complex structure which
would not occur in an axisymmetric galaxy (see Fig.~\ref{fig:all}
and \ref{fig:086}). The {\em global} appearance of a PVD can therefore be used
as a reliable diagnostic for the presence of a bar in an observed edge-on
disk. Specifically, the presence of a gap between the signatures of the
families of periodic orbits in the PVDs follows directly from the
non-homogeneous distribution of the orbits in a barred galaxy. The $x_1$ orbits
lead to a parallelogram-shaped feature in the PVDs which reaches very high
radial velocities with respect to the outer parts of the model when the bar is
seen end-on and rather low velocities when the bar is seen side-on. It
occupies all four quadrants of the PVDs, i.e.\ including the two forbidden
quadrants. This signature would dominate the structure of the PVD produced by
the stellar component of a barred spiral galaxy, and can be used as an
indicator of the viewing angle with respect to the bar in the edge-on
disk. When present, the $x_2$ orbits can also be used efficiently as a bar
diagnostic and behave similarly to the $x_1$ orbits in the PVDs. However, the
highest velocities are now reached when the bar is seen side-on and the
signature is spatially much more compact. The signature of the $x_2$ orbits
would dominate the structure of the PVD produced by the gaseous component of a
barred spiral.

The mass model we adopted had four free parameters, allowing to reproduce the
range of properties observed in real galaxies. Using the results of
\markcite{a92a}A92a, we analysed how the structures present in the PVDs vary
when the parameters of the model are changed. We predicted that the signatures
of the $x_1$ and $x_2$ periodic orbits are more extreme for high bar axial
ratios (except for the $x_2$ orbits), high Lagrangian radii, high central
densities, and/or for low bar quadrupole moments. In addition, the extent of
the $x_2$ orbits is reduced and can completely disappear when the bar axial
ratio and/or quadrupole moment are increased and when the central density
and/or Lagrangian radius are decreased. The shape and presence of the
signatures of the $x_1$ and $x_2$ familes of periodic orbits in a PVD can
therefore provide strong constraints on the mass distribution of an observed
galaxy.

We briefly discussed the application of the models to the interpretation of
real data. The major limitations of the models are the approximation of the
disk kinematics by that of periodic orbits, the treatment of the orbits as
``test particles'', and the ignorance of the production mechanism of the line
used in the observations. Nevertheless, the understanding of the traces of
individual orbits and of the signatures of orbit families in the PVDs will
prove indispensable in \markcite{ab99}Paper~II and \markcite{ab99b}Paper~III,
where, using hydrodynamical and $N$-body numerical simulations, we will
develop similar bar diagnostics addressing some of these limitations.
\acknowledgments
We thank K.\ C.\ Freeman and A.\ Kalnajs for comments on the manuscript, and
L.\ S.\ Sparke and A.\ Bosma for useful discussions in the early stages of
this work. M.\ B.\ acknowledges the support of an Australian DEETYA Overseas
Postgraduate Research Scholarship and a Canadian NSERC Postgraduate
Scholarship during the conduct of this research. M.\ B.\ would also like to
thank the Observatoire de Marseille for its hospitality and support during a
large part of this project. E.\ A.\ acknowledges support from the Newton
Institute during the final stages of this work.

\clearpage
%
%
%
\figcaption{(a) Characteristic diagram of model 001. Each point represents a
periodic orbit. The solid curves trace the extent of the various periodic orbit
families. Some families are identified. The zero-velocity curve (ZVC) which no
orbit can cross is
plotted with a dashed line. The dotted line represents circular orbits in the
azimuthally averaged mass distribution. (b) is simply an enlargement of
(a). (c) is the same as (a) but for model 086. The section of the $x_1$ family
plotted with a dot-dashed line represents orbits with loops.
\label{fig:chardiag}}
%
%
\figcaption{Basic properties of the $x_1$ and $x_2$ periodic orbits in model
001. (a) shows the axial ratios of the $x_1$ orbits as a function of their
semi-major (x$_{\scriptsize\mbox m}$, full line) and semi-minor
(y$_{\scriptsize\mbox m}$, dashed line) axes. (b) shows the velocities of the
$x_1$ orbits as they cross the minor axis of the bar (y$_{\scriptsize\mbox
m}$, full line), as well as the maximum velocities along the $x$-axis they
reach at any point (y$_{{\tiny\mbox V}{\tiny\mbox x}_{\tiny\mbox m}}$, dashed
line). (c) shows the velocities of the $x_1$ orbits as they cross the major
axis of the bar (x$_{\scriptsize\mbox m}$, full line), as well as the maximum
velocities along the $y$-axis they reach at any point (x$_{{\tiny\mbox
V}{\tiny\mbox y}_{\tiny\mbox m}}$, dashed line). (d)--(f) are analogous to
(a)--(c) but for the $x_2$ orbits. Note that both curves are superposed in
some plots.
\label{fig:a/bvxvy}}
%
%
\figcaption{(a) Lindblad precession frequencies in the azimuthally averaged
mass distribution of model 001. The various frequencies are identified on the
plot. The dotted curve shows the pattern speed $\Omega_p$ of the model.
(b) is the same as (a) but for model 086.
\label{fig:freq}}
%
%
\figcaption{The upper-left plot shows the face-on appearance (surface density)
of the orbits constituting the $x_1$ family of periodic orbits in model 001
(as selected in the text, \S~\ref{sec:orbitscal}). The other plots show
position-velocity diagrams (projected density
of material as a function of line-of-sight velocity and projected position
along the major axis) for the $x_1$ family when the galaxy is viewed
edge-on. The angle between the line-of-sight and the bar is indicated in the
top-left corner of each diagram, a viewing angle of $\psi=0\arcdeg$ indicating
that the bar is seen end-on (line-of-sight parallel to the bar) and a viewing
angle of $\psi=90\arcdeg$ indicating that the bar is seen side-on
(line-of-sight perpendicular to the bar)
\label{fig:x1}}
%
%
\figcaption{Same as Figure~\ref{fig:x1} but for the $x_2$ family of periodic
orbits in model 001.
\label{fig:x2}}
%
%
\figcaption{Same as Figure~\ref{fig:x1} but for the outer 2:1 family of periodic
orbits in model 001. Note that fewer viewing angles are displayed. 
\label{fig:out2_1d}}
%
%
\figcaption{Same as Figure~\ref{fig:x1} but for the $x_1$ and inner 4:1
families of periodic orbits in model 001. Note that fewer viewing angles are
displayed. Only the section of the inner 4:1
family lying outside the characteristic curve of the $x_1$ orbits in the
characteristic diagram has been considered.
\label{fig:4:1}}
%
%
\figcaption{Same as Figure~\ref{fig:x1} but for the $x_1$, $x_2$, LPO, and
outer 2:1 families of periodic orbits in model 001. Note that fewer viewing
angles are displayed. 
\label{fig:all}}
%
%
\figcaption{Same as Figure~\ref{fig:x1} but for the $x_1$ and outer 2:1
families of periodic orbits in model 086. Note that fewer viewing angles are
displayed.
\label{fig:086}}
%
%
%
\begin{figure}
\epsscale{0.45}
\plotone{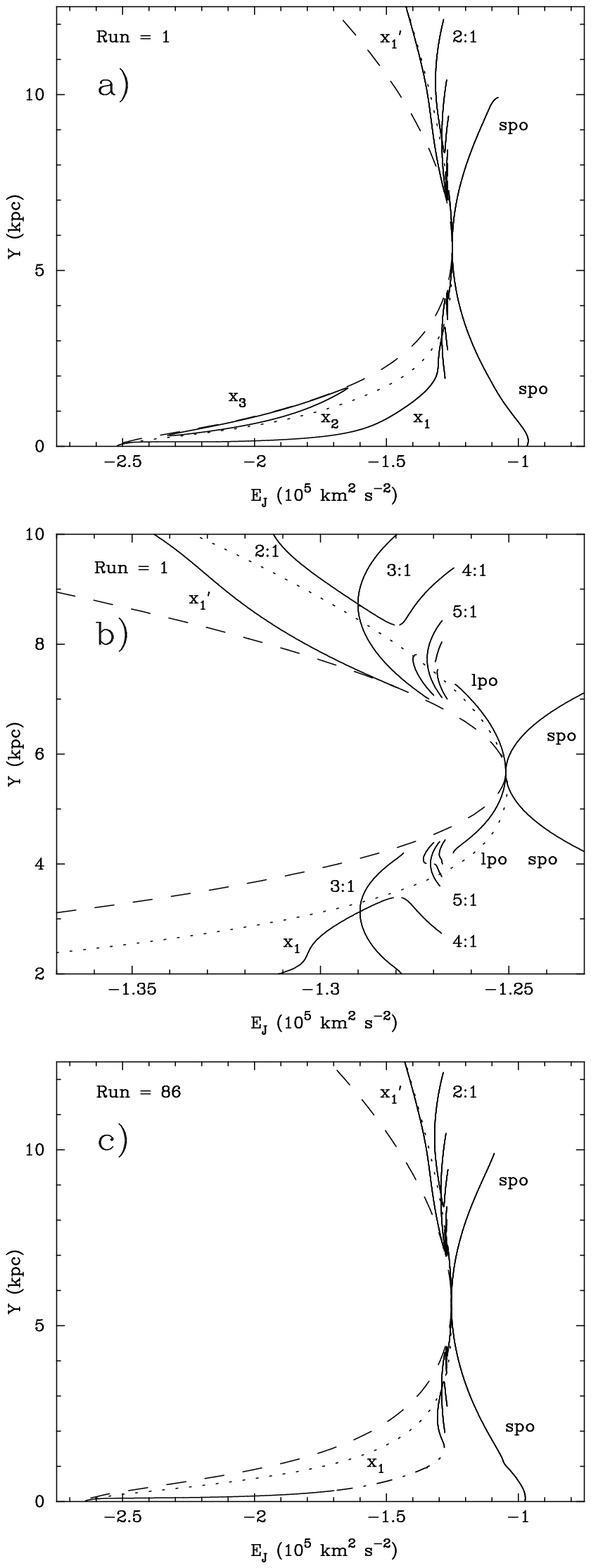}
\end{figure}
\epsscale{1.0}
\clearpage
%
%
\begin{figure}
\plotone{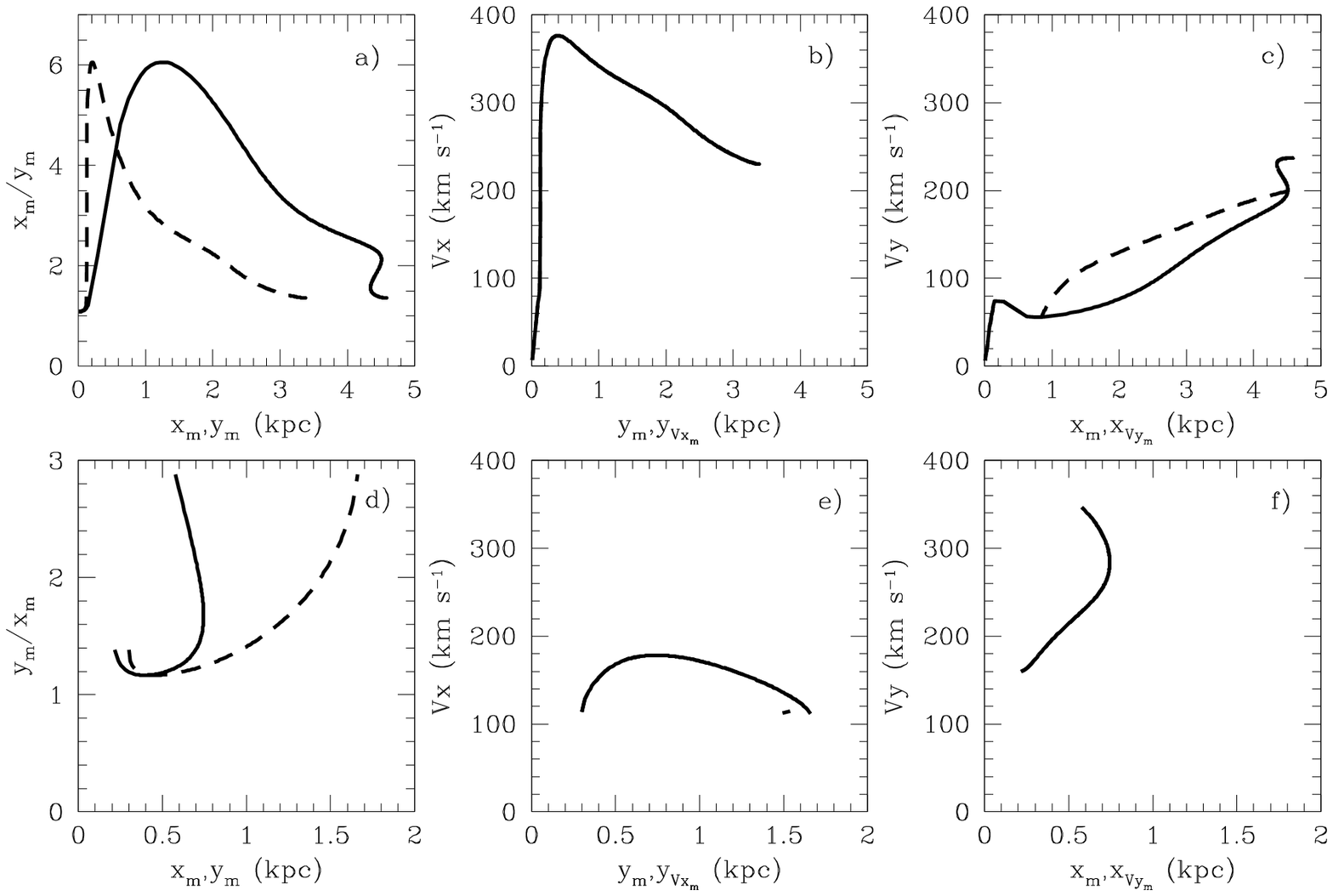}
\end{figure}
\clearpage
%
%
\begin{figure}
\epsscale{0.6}
\plotone{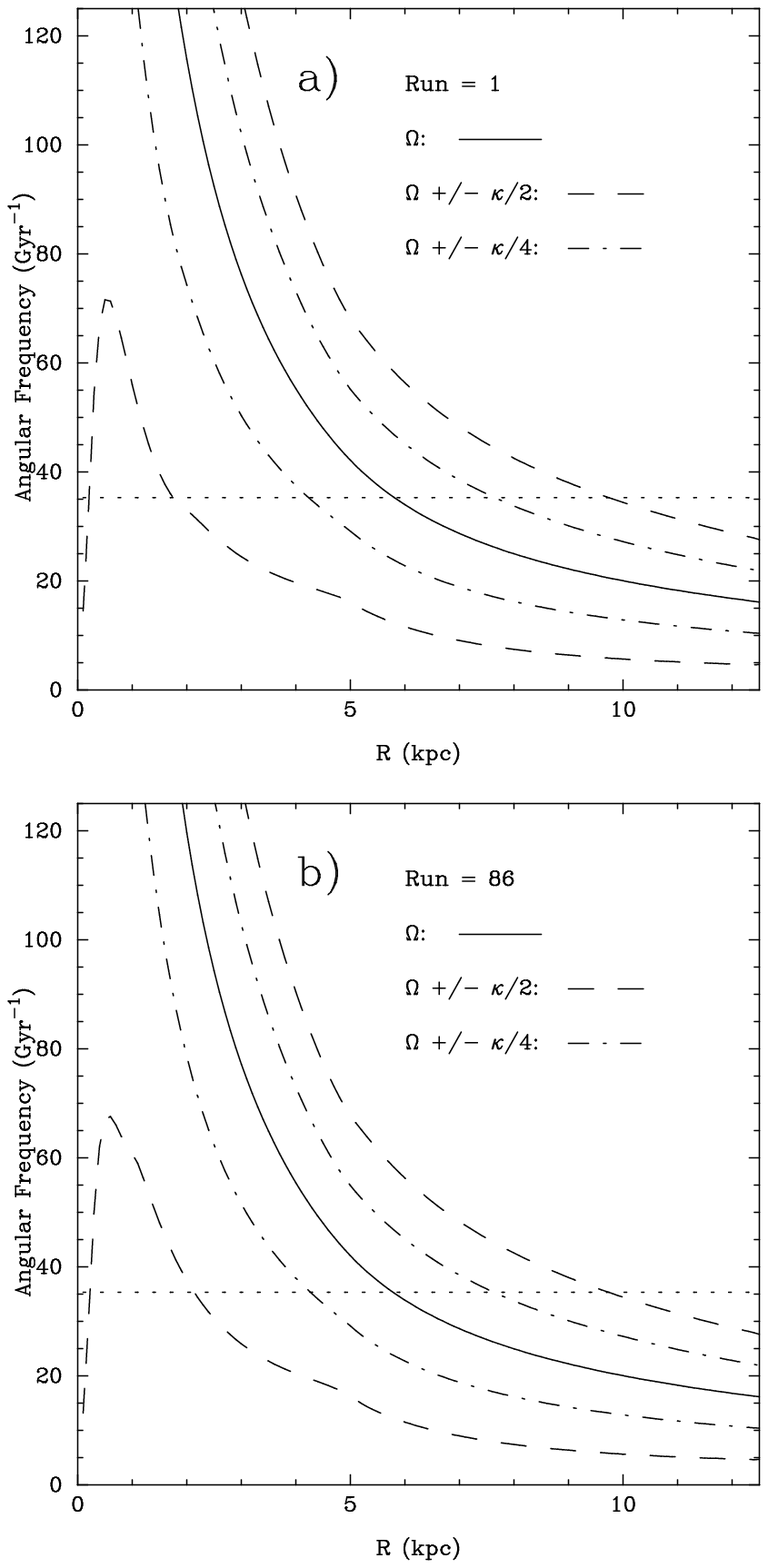}
\end{figure}
\epsscale{1.0}
\clearpage
%
%
\begin{figure}
\plotone{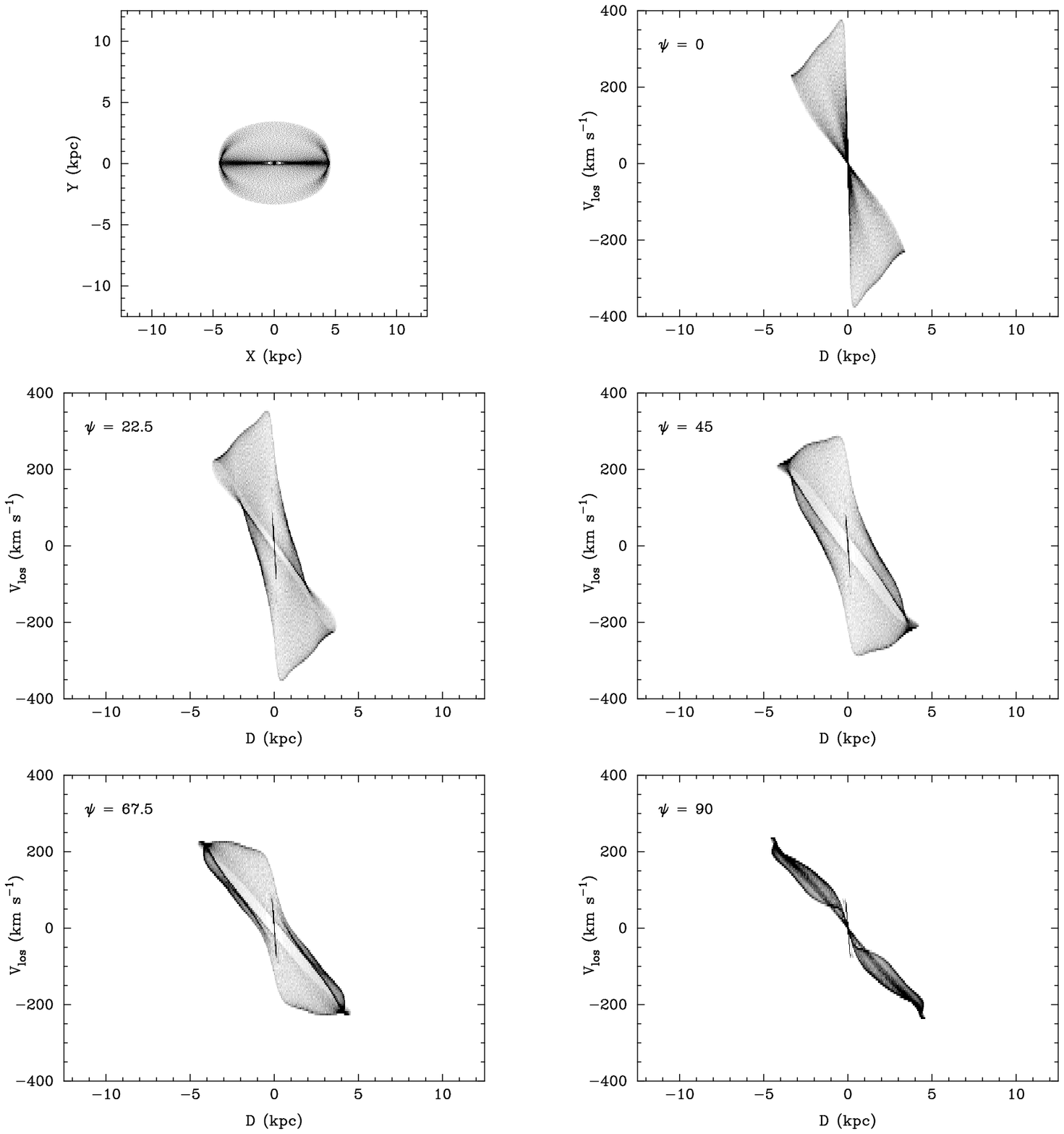}
\end{figure}
\clearpage
%
%
\begin{figure}
\plotone{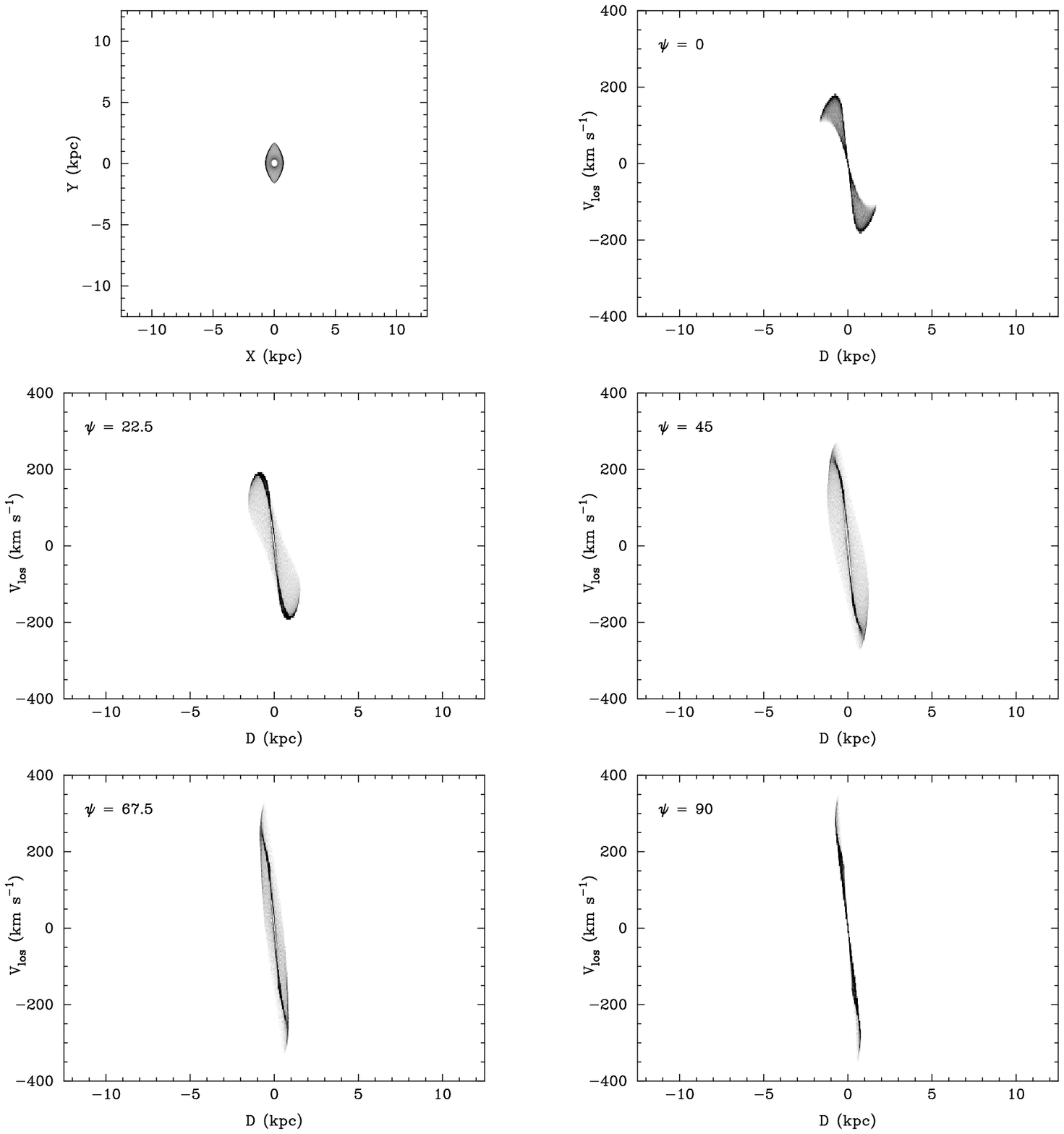}
\end{figure}
\clearpage
%
%
\begin{figure}
\plotone{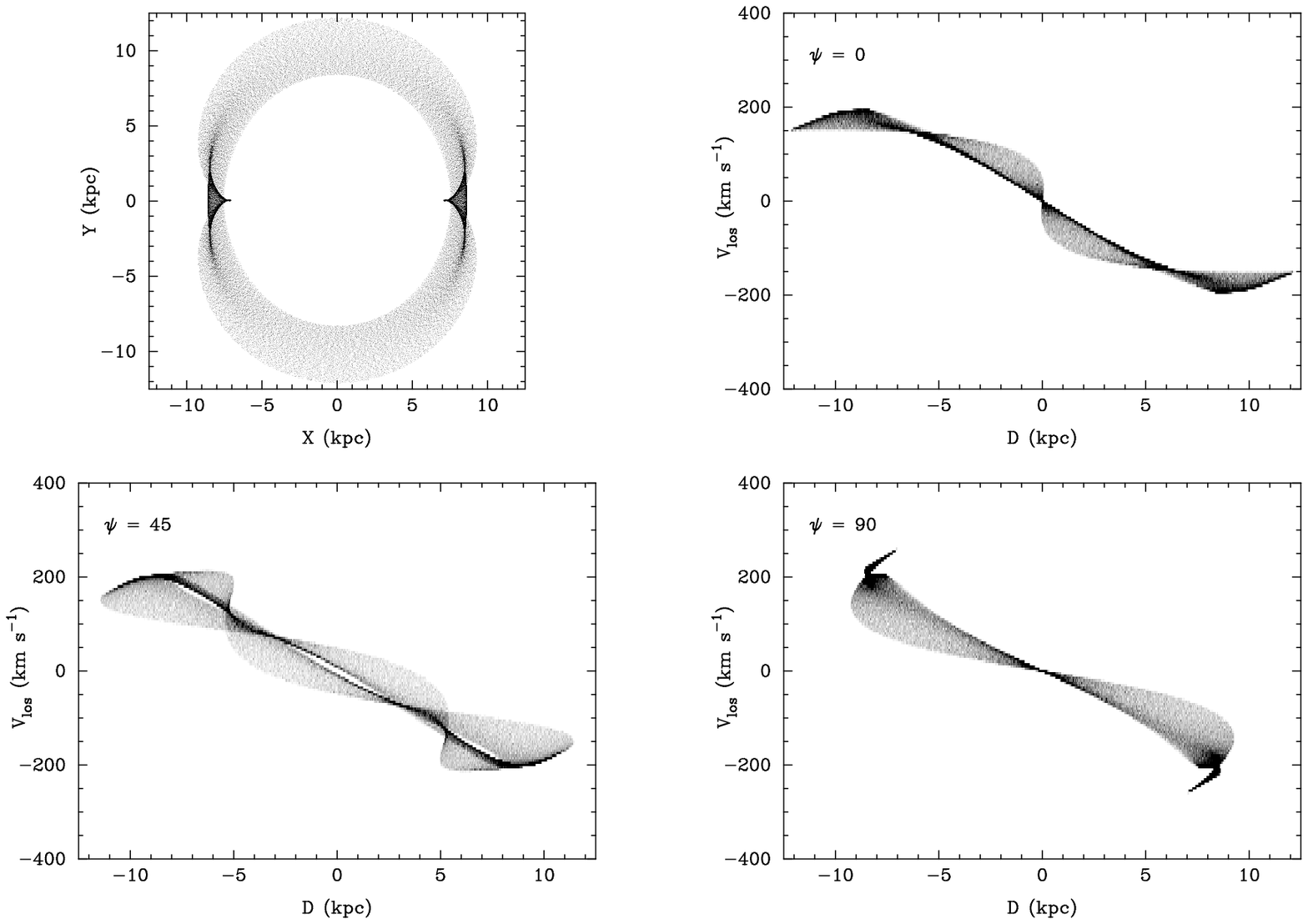}
\end{figure}
\clearpage
%
%
\begin{figure}
\plotone{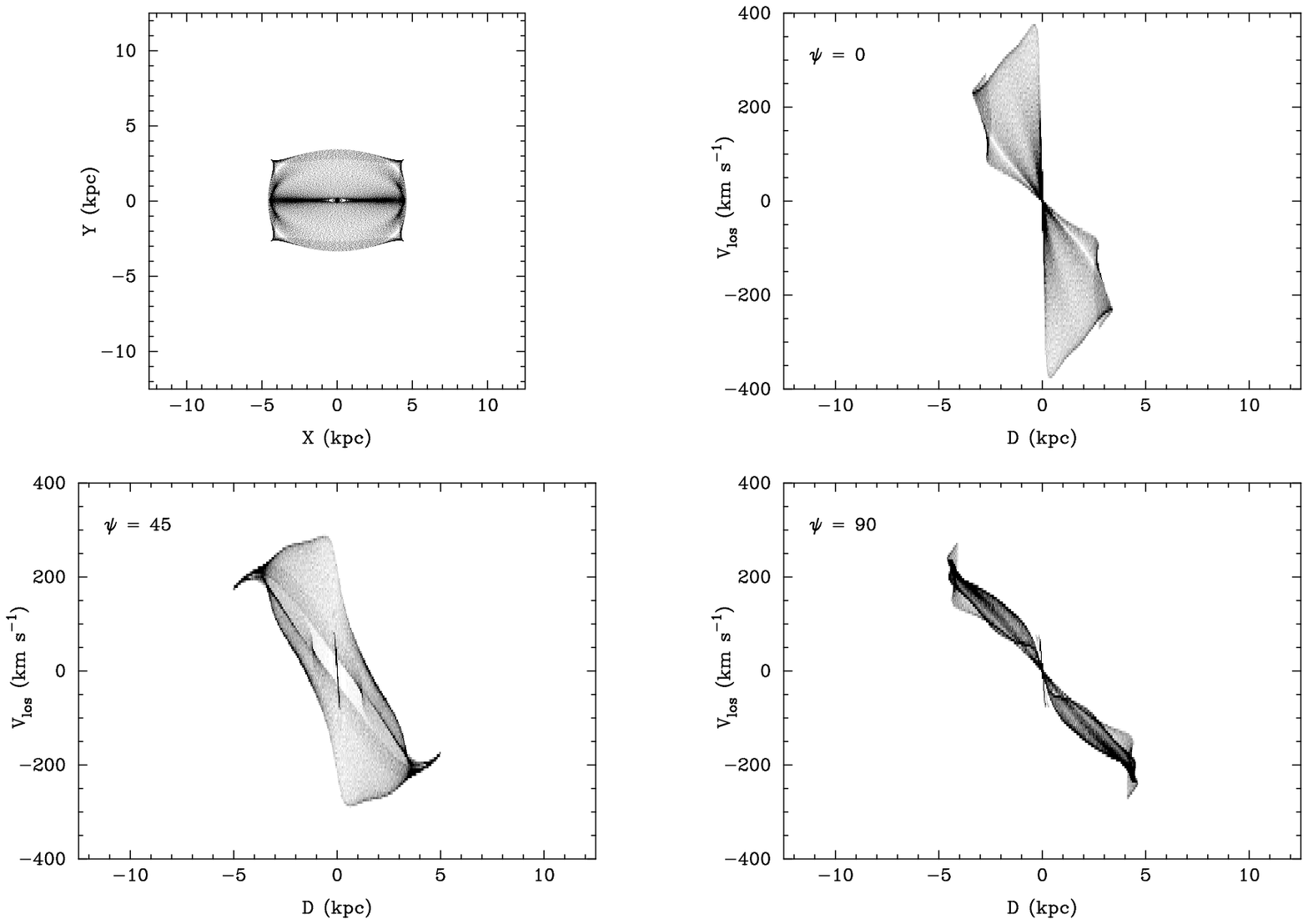}
\end{figure}
\clearpage
%
%
\begin{figure}
\plotone{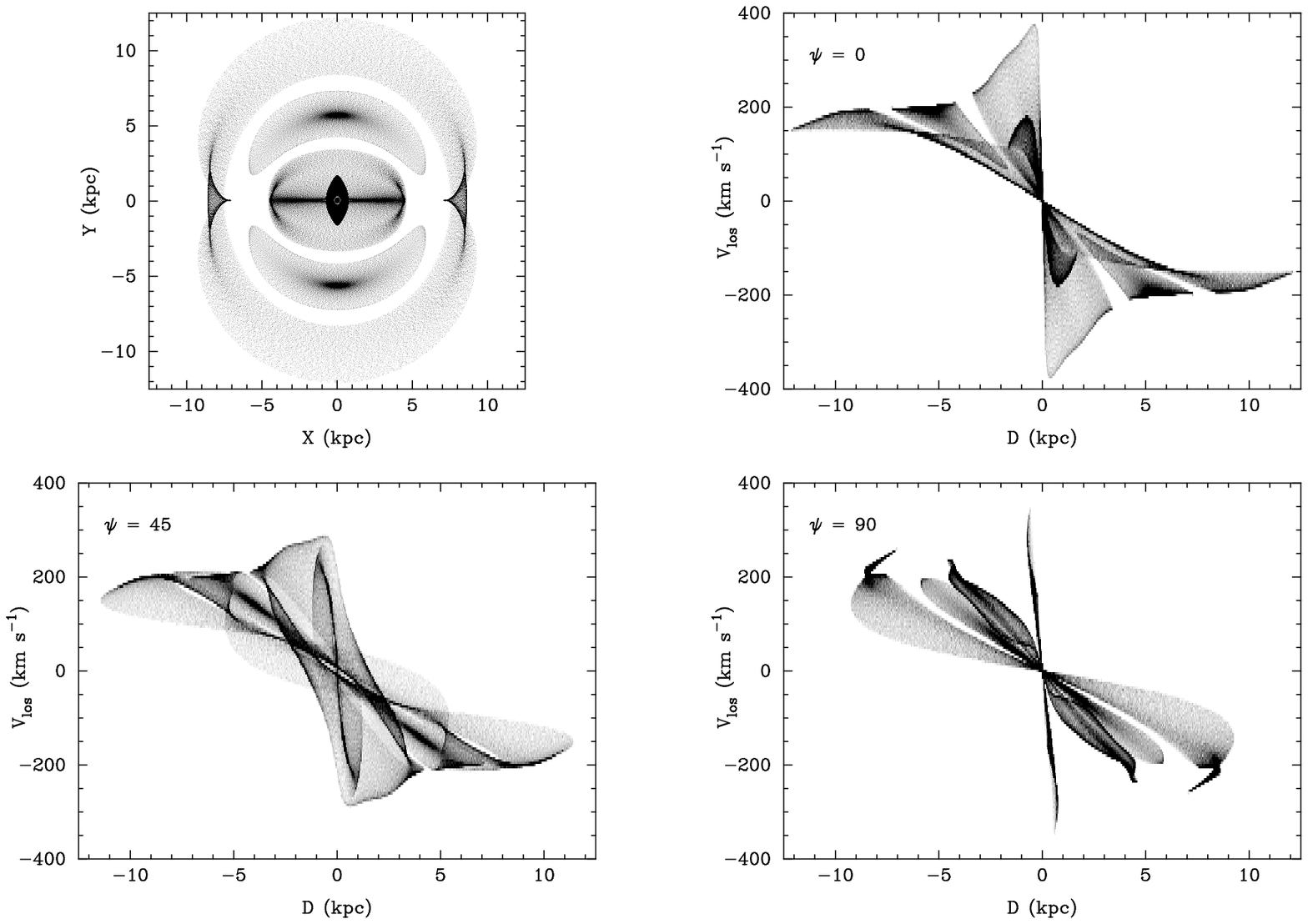}
\end{figure}
\clearpage
%
%
\begin{figure}
\plotone{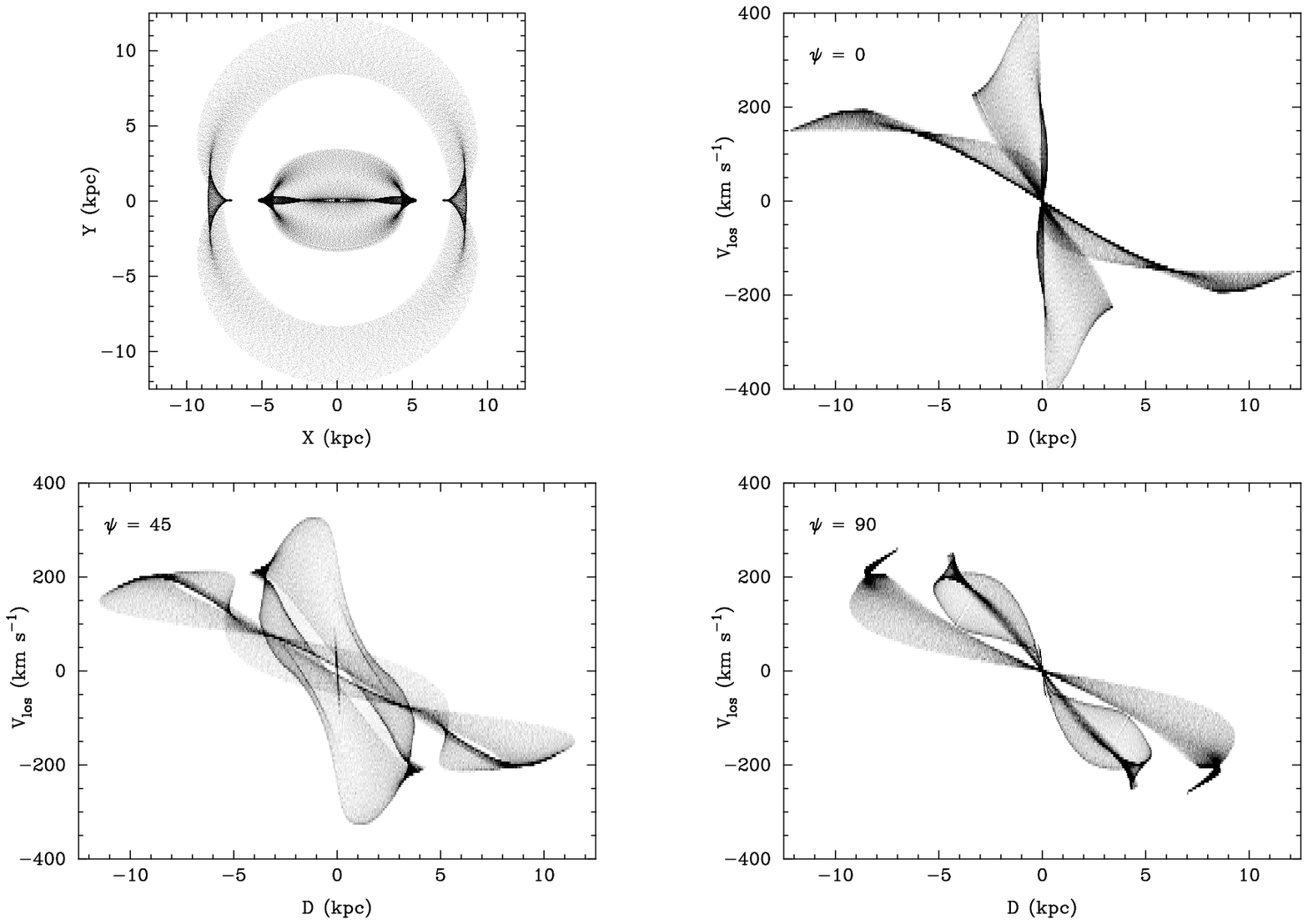}
\end{figure}
\end{document}